\documentclass[12pt,preprint]{aastex}

\def\lea{\mathrel{<\kern-1.0em\lower0.9ex\hbox{$\sim$}}}
\def\gea{\mathrel{>\kern-1.0em\lower0.9ex\hbox{$\sim$}}}

\slugcomment{Accepted for Publication in The Astronomical Journal}

\shorttitle{RR Lyraes in M33}
\shortauthors{Sarajedini et al.}

\begin{document}


\title{RR Lyrae Variables in M33. I. \\ Evidence For a Field Halo Population}

\author{Ata Sarajedini and M. K. Barker}
\affil{Department of Astronomy, University of Florida, Gainesville, FL 32611}

\author{Doug Geisler}
\affil{Grupo de Astronomia, Departamento de Fisica, Universidad de 
Concepci\'{o}n, Casilla 160-C, Concepci\'{o}n, Chile}

\author{Paul Harding}
\affil{Astronomy Department, Case Western Reserve University, 10900 
Euclid Avenue, Cleveland, OH 44106}

\author{Robert Schommer\footnote{deceased}}
\affil{Cerro Tololo Inter-American Observatory, National Optical Astronomy 
Observatories, Casilla 603, La Serena, Chile}



\begin{abstract}
We present observations of RR Lyrae variables in the Local Group late-type spiral
galaxy M33. Using the Advanced Camera for Surveys on the Hubble Space
Telescope, we have identified 64 ab-type RR Lyraes in M33. We have estimated
reddenings for these stars based on their minimum light $V-I$ colors and
metallicities based on their periods. From the distributions of these properties, we 
conclude that the RR Lyraes belong to two populations - one associated with the 
halo of M33 and the other with its disk. Given that RR Lyraes are 
produced by populations older than $\sim$10 Gyr, this suggests that not only
does the field halo of M33 contain an old component, but so does its disk.
This is one of the best pieces of evidence for the existence of a halo field
component in M33.
Using a relation between RR Lyrae absolute magnitude and metallicity
($M_V(RR) = 0.23[Fe/H] + 0.93$), we estimate a mean distance modulus of 
$\langle$$(m-M)_0$$\rangle$$ = 24.67 \pm 0.08$ for M33. This places M33
approximately 70 kpc beyond M31 in line-of-sight distance.
\end{abstract}


%% subject header.  Objects should be in the appropriate "individual"
%% headers (e.g. quasars: individual, stars: individual, etc.) with the
%% additional provision that the total number of headers, including each
%% individual object, not exceed six.  The \objectname{} macro, and its
%% alias \object{}, is used to mark each object.  The macro takes the object
%% name as its primary argument.  This name will appear in the paper
%% and serve as the link's anchor in the electronic edition if the name
%% is recognized by the data centers.  The macro also takes an optional
%% argument in parentheses in cases where the data center identification
%% differs from what is to be printed in the paper.


\section{Introduction}

The class of pulsating variables known as RR Lyraes have properties that
make them very useful for a number of astrophysical applications. 
The range of V-magnitudes they exhibit is fairly narrow and their light
curve shapes are distinctive (at least for the RRab types) making them easily
identifiable. Once identified, their presence is indicative of a stellar population
older than $\sim$10 Gyr; they can
be used as distance indicators; their minimum-light colors reveal the
line-of-sight extinction, and their pulsation properties (e.g. amplitudes and 
periods) are a reflection of their metallicities. 

Given their great utility, it is perhaps surprising that RR Lyraes have seen limited use
in the study of the stellar populations of M31 and M33 - the only other spiral
galaxies in the Local Group besides the Milky Way. Before the advent of the
refurbished Hubble Space Telescope, there was a serious dearth of such studies
because the faint magnitude of the RR Lyraes at the distances of M31 and M33
($V\sim$26) placed them near the magnitude limit of the largest ground-based telescopes.
An additional complication in the case of M33 has to do with its orientation 
on the sky which means halo stars are projected against the disk leading to
crowding and confusion problems.
%In the case of M33, its orientation means that image crowding is an 
%additional complication. 

One of the few ground-based studies focusing on the RR Lyraes in M31 was
that of Pritchet \& van den Bergh (1987). They observed 30 such stars in the 
halo of M31 using the CFH 3.6m telescope and an early generation RCA CCD.
They found a distance for M31 of $(m-M)_0 = 24.34 \pm 0.15$ based
on the mean magnitude of the RR Lyraes. This is slightly closer but still consistent
with the (generally accepted) value of $(m-M)_0 = 24.47 \pm 0.03$, which
represents the weighted mean of seven different recent determinations 
(Rich et al. 2005). 

Ground-based observations of RR Lyraes in M33 were presented by Pritchet (1988).
He describes {\it preliminary} results based on 7 of the best candidate RR Lyraes
and computes a distance of $(m-M)_0 = 24.45 \pm 0.2$. To put this number in
perspective, we note that Galleti,  Bellazzini, \& Ferraro (2004) have collected 
distance estimates to 
M33 using a variety of techniques. The Pritchet (1988) value is on the low side of these
various distances. However, the results of Pritchet (1988) remain
preliminary and unpublished in the refereed literature. 

With the advent of the Hubble Space Telescope (HST), the study of HB stars, 
and in particular RR Lyraes, in M31 and M33 has become much more tractable. 
The higher spatial resolution of HST has made it
possible to reliably photometer stars in the densest regions of these galaxies
(e.g. the bulge of M31, Sarajedini \& Jablonka 2005) as well as to magnitudes
fainter than the main sequence turnoff in M31 (Brown et al.
2003). The time coverage available from the latter set of observations was
exploited by Brown et al. (2004) to study 55 RR Lyraes in a $\sim$11 square
arcmin  region in the halo of M31. 

In contrast to M31, there is a paucity of RR Lyrae studies of M33, both from the
ground and from space. As a result, we herein describe the results of such a 
study, which was originally aimed at obtaining deep photometry of two star
clusters in M33. As such, although the data are not ideal for an RR Lyrae search, 
they represent the first space-based study of RR Lyraes in M33 and provide 
important information about this population. 
Sections 2 and 3 describe the observations and reductions, respectively.
The characterization of the variable stars in also described in Sec. 3. Information 
that the RR Lyraes provide about the reddening to M33, 
the metallicity distribution, and the distance of M33 are
all discussed in Sec. 4. We discuss our results in Sec. 5.
Finally, the conclusions are given in Sec. 6.

\section{Observations}

The images used in the present study were obtained with HST 
using the Advanced Camera for Surveys instrument (ACS) 
as part of our GO-9873 program. As shown in Table 1 and Fig. 1, two fields were observed in 
M33 - one in the vicinity of the star cluster M9 and the other including the clusters 
U49 and H10. Each exposure indicated in Table 1 was CR-SPLIT resulting in a 
total of 24 images for each field, 8 in the F606W filter
and 16 in the F814W filter.  A box-shaped dither pattern was applied with offsets of 
$\sim$0.25" between each CR-SPLIT exposure. The temporal coverage of each field 
is $\sim$2.2 days.

\section{Reductions}

\subsection{Construction of Point Spread Functions}

The program frames of M33 are extremely crowded making the 
construction of point spread functions (PSFs) for use in crowded-field 
photometry quite challenging. In order to more easily construct high S/N 
PSFs, we have relied on archival images of the globular cluster 47 Tuc 
obtained as part of three ACS programs: 9656 (PI: De Marchi), 10048 (PI: Mack),
and 10375 (PI: Mack). These are relatively uncrowded fields located 
$\sim$6.7 arcmin from the center of 47 Tuc; thirteen such images are 
available in the F606W filter and the same number in F814W. 

Each WFC1 and WFC2 FLT image was multiplied by the 
corresponding geometric correction frame and used to generate 
a PSF via the following procedure. We used the DAOPHOT/ALLSTAR 
suite of software (Stetson 1987) to find stars on each image and produce 
small-aperture photometry for them. We then picked 1000 candidate PSF 
stars. After deleting those with bad pixels nearby, we subtracted the 
neighboring stars surrounding these PSF stars. The resultant list of between 
500 and 700 stars was used to produce a PSF for each of the 13 images
in each filter. The shape of the PSF was made to vary quadratically with 
position on the frame. For each independently derived PSF, the ADDSTAR 
task in DAOPHOT was used to 
place 3,240 faint stars  on a zero image using the same
fixed grid pattern (40 x 81 stars). 
The resultant 'stack' of 3,240 stars composed of 13 separate and independent PSFs 
was used to produce a single high S/N PSF for F606W and F814W.

\subsection{Photometry of Program Frames}

Each of the program FLT images was multiplied by the corresponding 
geometric correction image; the bad pixel masks (i.e. data quality files) 
were applied and the resultant frames of each ACS chip (WFC1 and WFC2)
were photometered with the DAOPHOT/ALLSTAR/ALLFRAME routines
(Stetson 1994). We followed
the standard procedure for producing crowded-field photometry using
these programs (Sarajedini et al. 2000)
except that we employed the PSFs described above instead of deriving 
PSFs from the frames themselves. 

To summarize, the process involves producing a coordinate
transformation between all of the frames (24 in this case). This transformation 
is used to produce a median combined image using all of the frames, which is 
then run through 
ALLSTAR/FIND/PHOTOMETRY three times to produce a master coordinate list of 
detected stellar profiles. The coordinate transformation along with the 
master list and the PSFs are inputs to ALLFRAME which yields profile-fitting 
magnitudes and positions for stars on each of the 24 frames. 

Stars appearing on all of the frames (8 in F606W and 16 in F814W) were matched 
to form mean instrumental magnitudes. These have been standardized 
using a two step process. First aperture corrections were derived to 
correct the PSF mags to an aperture size of 7 pixels where the charge 
transfer efficiency (CTE) corrections of Reiss (2003) were applied. The 
magnitudes were then corrected to an infinite radius aperture and
transformed to V and I using the relations published  by Sirianni et al. 
(2005). They quote magnitude differences between a 10 pixel radius
aperture and infinity (a radius of 5.5 arcsec) for {\it drizzled} (DRZ) ACS images.
However, because we have photometered the FLT images, it is important
to be sure that the magnitude differences between a 10 pixel radius
aperture and infinity are the same as those measured for DRZ frames.
Therefore, using three F606W and three F814W FLT images of 47 Tuc,
the same ones  used to 
construct our PSFs, we find mean differences of 0.019$\pm$ 0.011 mag
and 0.031$\pm$0.019 (Us -- Sirianni) between our aperture corrections
and those of Sirianni et al. (2005). Because these differences are not
statistically significant, we can safely apply the Sirianni et al. values
to our photometry. Based on the discussion in Sec. 8.3 of Sirianni et al.
(2005), we adopt an error of $\pm$0.05 mag in our photometric zeropoint.

\subsection{Characterization of Variable Stars}

In order to study the variable stars most effectively, we used DAOMASTER 
to match the stars with measurements on all 24 frames. Stars with frame-to-frame
standard deviations greater than 0.2 mag were selected as candidate variable 
stars for further study. This resulted in $\sim$400 candidates in each WFC
chip. For comparison, the typical frame-to-frame standard deviation
for non-variables stars at the same magnitude as the RR Lyraes is $\sim$0.08 mag.
The time-series photometry for these stars was then inspected by eye
to gauge the quality of the variation. Those light curves deemed to be of
sufficiently high quality were analyzed further via the template-light-curve fitting 
procedure described by Layden \& Sarajedini (2000; Pritzl et al. 2002; Mackey
\& Gilmore 2003). The code written by Andy Layden and available from
his web page
\footnote{http://physics.bgsu.edu/~layden/ASTRO/DATA/EXPORT/progs.htm} 
uses 10 template light curves
(6 RRab type variables, 2 RRc types, an eclipsing binary, and an algol binary).
It searches over a given period range and calculates the $\chi^2$ difference between
the template and the data. The period that yields the minimum  $\chi^2$ with a
given light curve is then fit to that light curve to determine the properties of the
variable. For the purposes of the present paper, we are only concerned
with the RR Lyrae-type variables. Because the light curves are better sampled
in the I-band, the analysis was initially performed on these magnitudes using
the V-band data to help constrain the phase. Once an 
acceptable period and phase were determined from the I-band data, these
were used to fit the
light curve templates to the V-band observations. In one case, the
V-band data do not provide sufficient phase coverage for such a fit 
(U49-WFC1-131455). Based on our experience with the application of
the above-mentioned technique to the observed data points, we estimate 
an error of $\pm$0.01 day in the derived periods. That is to say, most of the time
a change of $\pm$0.01 day in the optimum period would yield a phased
light curve of  similar quality, but a change of $\pm$0.02 day would frequently
yield a light curve of significantly worse quality. The error is dominated by
the particular time sequence of the observations; thus, it is the same for all
of the RR Lyrae stars. It should be 
noted that we checked and re-checked the period determinations of all the 
RR Lyraes to be certain of their robustness. Given the quality of our data, we
are confident that we have derived the best periods possible.

\section{Results}

The procedure described above yielded 72 RR Lyrae variables in the two
ACS fields considered here. Of these, the vast majority (64) are of the 
RRab type. Table 2 gives the relevant information for each variable.
These include the Right Ascension, Declination, intensity weighted
mean V and I magnitudes, magnitude weighted  mean V--I color, the period in 
days, the I-band amplitude, and the type of RR Lyrae.
Figures 2 through 11 show the light curves of the variables discovered in the present study
designated by the field (M9 or U49) and CCD chip (WFC1 or WFC2) in which
they are found. 
%They are arranged in order of magnitude with the brightest stars
%in each chip shown first. 
It is clear from the light curves that, in some cases, the V-band amplitudes are
especially problematic. These should generally be $\sim$1.6 times greater
than their I-band counterparts (Liu \& Janes 1990, Table 6). This is not the 
case for a number of RR Lyraes in our sample principally because the phase
coverage is not sufficient in the V-band. 

Figure 12 illustrates the location of the RR Lyraes in the $(V,V-I)$ 
color-magnitude diagram
(CMD) along with the photometry for the WFC1 chip of the M9 field as a
reference. We see that most of the RR Lyraes are located where
we expect to see such stars. There is a clear dispersion in their locations 
in the direction of higher reddening/extinction in the CMD. This is our first indication
that not all of the RR Lyraes are located on the near side of M33.

The blue edge of the RR Lyrae instability strip has a color
of $(V-I)_0 = 0.28\pm 0.02$ (Mackey \& Gilmore 2003), which given the 
line-of-sight reddening toward M33
of $E(V-I) = 0.06\pm 0.02$ (Sarajedini et al. 2000), translates to $(V-I) = 0.34$. 
Thus, we expect no RR Lyrae stars bluer than this color, yet Fig. 12 shows 4 stars
(M9-WFC2-109239, -112403; U49-WFC1-72044, -88219) that
are significantly bluer than $(V-I) = 0.34$.  The latter two stars are c-type RR Lyraes,
which are generally bluer than ab-types. From Fig. 11, it appears that the
color of M9-WFC2-109239 is indeed close to zero. In the case of 
M9-WFC2-112403, the V-band amplitude is underestimated, which could contribute
to an overestimate in its inferred V magnitude, thus making the color too blue. 
Another star in an anomalous location in the CMD is
M9-WFC1-21307 located at V$\sim$24.6 and (V-I)$\sim$0.33. Again, the
light curve appears to be normal, but the magnitude of the variable is too
bright by $\sim$0.5 mag to be in the canonical RR Lyrae region. This is also
the only RR Lyrae in our sample that is located in the vicinity of a star cluster 
- that being M9.  We suspect the lack of RR Lyrae stars in these 
clusters is due to their extreme crowding as well as the possibility 
(Sarajedini et al. 2000) that they are younger than $\sim$10 Gyr.

Figure 13 shows the period-amplitude diagram for the RR Lyrae variables.
The open circles are the ab-type variables, and the filled circles are those
that exhibit sinusoidal light curves indicative of being c-type RR Lyraes.
The solid lines are the loci of ab-type RR Lyraes in Oosterhoff I (left) and II (right)
globular clusters taken from Clement (2000). The V-band amplitudes ($Amp_V$) of
Clement (2000) are converted to I-band values ($Amp_I$) using 
$Amp_V=1.6Amp_I$ determined from Table 6 of Liu \& Janes (1990). 

A close inspection of Fig. 13 suggests that 5 of the putative c-type RR Lyraes
(U49-WFC1-72044, -88219, -102459, -109540, M9-WFC1-60843) occupy 
a region that is representative of ab-type RR Lyraes. This suggests that these
stars are either misclassified ab-type RR Lyraes or that they are an entirely different
type of pulsating variable. In addition, c-type variable U49-WFC1-95229 has
a period (Log P $\sim$ --0.6) that is much shorter than any of the RR Lyraes 
probably indicating that it is not an RR Lyrae at all. 
Because of the ambiguity with regard to the nature of the c-type variables,
we will focus only on the RRab variables in the subsequent analysis.

Figure 14 illustrates the period distribution for the RRab type variables.
This is compared with RRab stars in the M31 halo (dashed line)
from Brown et al. (2004) and
in the Galactic globular cluster M3 (dotted line), which is classified as Oosterhoff I 
(Clement 2000; Brown et al. 2004). The M33 histogram exhibits a primary peak at
Log P $\sim$ --0.27 corresponding to a period of 0.54 days, which is 
consistent with the mean of 0.55 days characteristic
of Oosterhoff I clusters like M3. This is also apparent in the period-amplitude
diagram (Fig. 13). The M31 distribution shares a peak with the RR Lyraes in M3, but
exhibits a tail to longer periods.

Unlike M31 and M3, the RRab star distribution in M33 exhibits an additional 
portion that extends to periods as short as $\sim$0.3d. Dividing the RRab sample
at 0.44d (Log P = --0.36), we find that 38\% of the RRab stars have periods
shorter than this value. In contrast, this percentage is only $\sim$17\% among the solar
neighborhood RR Lyraes represented in the Northern Sky Variability Survey
(Kinemuchi et al. 2005). As we will see in Sec. 4.2, because period and
metallicity are correlated for RRab stars, these short period RR Lyraes are
most likely an old metal-rich component in the disk of M33. These would necessarily
not be seen in a globular cluster like M3 nor the halo of M31.

%The thick dashed vertical
%line in Fig. 14 represents the shortest period (i.e. most metal-rich, $[Fe/H] \sim 0$) 
%RRab variables in the Layden (2005, private communication) database representing
%solar neighborhood RR Lyraes. If we take Fig. 14 at face value, 
%M33 appears to harbor RR Lyraes that have significantly super-solar 
%metallicities. We return to a more detailed discussion of these stars in Sec. 4.2. 
%For the moment, we note that the RRab stars with periods less than 0.44d 
%(Log P = --0.36) account for 38\% of the total number of such stars in our
%M33 fields. In contrast, this percentage is only $\sim$17\% in the solar
%neighborhood RR Lyraes represented in the Northern Sky Variability Survey
%(Kinemuchi et al. 2005).

\subsection{RR Lyrae Reddenings}

The minimum light color of RRab stars can be used to estimate their line-of-sight
reddenings. This is based on a concept originally developed by Sturch (1966), expanded
on by Mateo et al. (1995), and
most-recently refined by Guldenschuh et al. (2005). The latter work concludes that
the intrinsic minimum light color of RRab variables is $(V-I)_{0,min} = 0.58 \pm 0.02$
with very little dependence on period or metallicity for periods between 0.39d and 0.7d
and metallicities in the range $-3\lea[Fe/H]\lea0$.

Of the 64 RRab types in our M33 dataset, we were able to fit V-band light curves
to all of them except U49-WFC1-131455, which does not possess adequate phase
coverage in the V-band. From these light curves, we determine the minimum-light color and use
it to calculate the line-of-sight foreground reddening for each RRab star. The 
distribution of these reddenings
is shown by the binned histogram in Fig. 15. For reference, note that the line-of-sight 
reddening to M33
is $E(V-I) = 0.06 \pm 0.02$ (Sarajedini et al. 2000). Therefore, we see stars on
the near side of M33, reddened only by dust in the Milky Way ($E(V-I) < 0.1$), 
that are likely to be members of the M33 halo population. We also see 
RR Lyraes that are reddened by an amount that is in 
excess of the line-of-sight value suggesting that they are either in the disk
of M33 or halo field stars on the far side. The large range of reddening illustrated
in Fig. 15 is supported by the presence of a sloping red clump (RC) in the CMD of Fig. 12,
wherein the RC slope is consistent with the reddening vector. The reddening
range is also consistent with and therefore supported by the results of 
Boissier et al. (2004) and Massey et al. (1985).

There are 18 stars in Fig. 15 that have negative reddenings, which is clearly unphysical.
However, the overall distribution is strongly peaked around zero.
Inspection of their light curves shows no clear pattern that would explain this
behavior, but, since the reddening distribution is not symmetric around zero, the 
negative portion of the distribution can be interpreted in terms of the 
errors in our analysis. Fitting a Gaussian
function (dotted lines in Fig. 15) to the negative portion yields 
$\sigma$ = 0.16 mag in $E(V-I)$, while a
similar fit to the positive values gives $\sigma$ = 0.34 mag. Subtracting these
in quadrature provides a measure of the extinction in the disk of M33 of
$\sigma$ = 0.30 mag in $E(V-I)$. Utilizing the Sirianni et al. (2005) relations
for extinction in the HST/ACS filters
gives a 1-$\sigma$ extinction of $A_{F606W}$ = 0.62 mag and 
$A_{F814W}$ = 0.39 mag.

To place these results in the context of other galaxies, we look to the paper
by Holwerda et al. (2005), who investigate the radial extinction properties
of nearby spiral galaxy disks using counts of distant galaxies. Their Fig. 4 presents
a plot of I-band extinction ($A_I$) as a function of radial location in the 
galaxy expressed in terms of $R_{25}$, which is half the $D_{25}$ diameter
given in the Third Reference Catalog of Bright Galaxies (RC3). For M33,
this $R_{25}$ radius is 35.4 arcmin while the mean deprojected radius
of the RR Lyraes is 13.3 arcmin (both fields are at very similar radial
distances). Looking at Fig. 4 of Holwerda et al. (2005)
at a radius of $R/R_{25}$ = 0.38, we see that an extinction value ranging
from $A_{F814W}$ = 0.39 mag (1-$\sigma$) to $A_{F814W}$ = 0.78 mag
(2-$\sigma$) is consistent with those of the spiral galaxies in 
Holwerda et al.'s sample.

\subsection{Metallicities}

A number of authors have studied the relationship between the metal abundance
of RRab variables and their pulsation periods (e.g. Sandage 1993; Layden 1995). 
It is well known that as the metallicity of RRab's increases, their periods
decrease. We can exploit this fact to investigate the metal abundances of the
M33 RR Lyraes studied herein.

Using the data of Layden (2005, private communication) for 132 Galactic RR Lyraes
in the solar neighborhood,
we can establish a relation between period and metal abundance. Figure 16
shows these data along with a best-fit line derived by performing a least-squares
fit using Log P as the independent variable and then with [Fe/H] as the independent
variable and combining the results via the OLS Bisector method (Isobe et al.
1990). We find

\begin{eqnarray}
[Fe/H] = -3.43 - 7.82~Log~P_{ab}~~~~~~~~~~~~~~rms = 0.45~dex.
\end{eqnarray}

\noindent This compares favorably with 
$[Fe/H] = -4.34 - 10.87~Log~$$\langle$$P_{ab}$$\rangle$ determined by
Brown et al. (2004) based on the findings of Sandage (1993). Both equations
are on the Zinn \& West (1984) abundance scale. More importantly, the
equation derived herein is valid for a larger range of metallicities 
($-2.5$$\lea$$[Fe/H]$$\leq$0.0).

Figure 17 shows the distribution of M33 RRab metallicities computed using
this equation.  The most dominant feature is a strong peak at 
$[Fe/H] \sim -1.3$. 
In addition, there are RR Lyraes with metallicities greater than the solar
value. These correspond to the stars with periods shorter than $\sim$0.35 days
mentioned above. Since our metallicity calibration does not extend beyond
$[Fe/H] \sim 0.0$, these super-solar abundances should be considered
with caution and must therefore receive less weight. 

In order to quantitatively assess the significance of the apparent bimodality
in the metallicity distribution, we utilize the KMM mixture-modeling algorithm
(McLachlan \& Basford 1988). This algorithm objectively partitions a
dataset into two (or more) components and then assesses the improvement
of the two-group fit to the one-group fit using the likelihood ratio
test statistic (see Ashman, Bird \& Zepf 1994 for a discussion of this
algorithm in the context of astronomical applications).
We find that 
the KMM algorithm rejects the unimodal hypothesis in favor of the
bimodal one at a confidence level of $> 99.9\%$, irrespective of whether the 
stars with supra-solar metallicities are assigned their nominal values or are
simply assigned a metallicity of 0. This confirms the reality of the two 
metallicity populations that appear visually.

Attributing the root-mean-square residual of the fit shown
in Fig. 16 (0.45 dex) as the error in each individual abundance, we can construct
a generalized histogram of the metallicity values. This is shown as the dashed
curve in Fig. 17. For reference, we note that at the mean deprojected
galactocentric distance of the RR Lyraes 
(3.36$\pm$0.05 kpc), the metal abundance of the red giant branch stars
in the  M33 disk is 
$[Fe/H] = -0.72 \pm 0.03$; this is  based on the relation published by Kim et al. (2002)
of $[Fe/H] = -0.05 (\pm 0.01) R_{deproj}(kpc) - 0.55 (\pm 0.03)$.
In addition, the mean metallicity of halo globular
clusters in M33 is $[Fe/H] = -1.27 \pm 0.11$ (Sarajedini et al. 2000).
The solid line in Fig. 17 shows two Gaussian distributions, one with a peak at 
$[Fe/H] = -0.7$ (disk) and the other at $[Fe/H] = -1.3$ (halo) fitted to the 
generalized histogram (dashed line).
The fact that the Gaussian fits are consistent with the observed metallicity
distribution of the RR Lyraes suggests that the metal-rich population can be
attributed to the disk of M33 while the more metal-poor population belongs
to the halo. This confirms what we concluded 
from the distribution of RR Lyrae reddenings, namely that our RR Lyrae sample
is composed of both halo and disk populations.

\subsection{Distance of M33}

As noted in Sec. 1, we can make use of the RR Lyrae stars to determine the
distance to M33. We will not use stars with super-solar abundances
because our Log P $\sim$ [Fe/H] calibration is uncertain for these stars.
The mean V magnitude of the 43 RRab stars with positive reddening values and
metallicities less than solar is
$\langle$$V(RR)$$\rangle$$ = 25.92 \pm 0.05$ (ran) $\pm0.05$ (sys), where the 
random error is the standard error of the mean and the systematic error represents
the uncertainty in the photometric zeropoint. Applying the reddenings to
the apparent magnitudes using $A_{F606W} = 2.1 E(V-I)$ gives a mean
intrinsic V magnitude of $\langle$$V_0(RR)$$\rangle$$ = 25.34 \pm 0.07$,
where we have added the random and systematic errors in quadrature.

The next step involves adoption of metallicities for each RR Lyrae star. 
This has already been discussed above
where we presented our equation relating the RR Lyrae's pulsation period with
its metal abundance. To calculate the absolute magnitude from the metallicity, we rely 
upon the relation $M_V(RR) = 0.23[Fe/H] + 0.93$ derived by
Chaboyer (1999). In this case, the mean absolute
distance modulus for the 43 RRab stars with positive reddenings and subsolar 
metallicities turns out to be
$\langle$$(m-M)_0$$\rangle$$ = 24.67 \pm 0.08$. This value for the distance 
modulus is in very good agreement with the results of
Galleti et al. (2004) who find an average distance of 
$\langle$$(m-M)_0$$\rangle$$ = 24.69 \pm 0.15$ based on 19 independent
estimates. 

% Excluding all stars with mets greater than [Fe/H] = 0.0
% $\langle$$V(RR)$$\rangle$$ = 25.918 \pm 0.05$ (ran) $\pm0.05$ (sys)
%$\langle$$V_0(RR)$$\rangle$$ = 25.342 \pm 0.07$
%$\langle$$(m-M)_0$$\rangle$$ = 24.67 \pm 0.09$

Since our mean I-band magnitudes are generally better determined than 
those in the V-band, we can also estimate the distance of M33 using the I-band
data. Doing this, we find 
$\langle$$I(RR)$$\rangle$$ = 25.11 \pm 0.03$ (ran) $\pm0.05$ (sys) for
the 43 RR Lyraes in our subsample. Applying the reddenings for
each individual star yields $\langle$$I_0(RR)$$\rangle$$ = 24.73 \pm 0.06$.
Employing once again the Chaboyer (1999) equation relating RR Lyrae absolute
magnitude and metallcity, we find $\langle$$M_V(RR)$$\rangle$$ = 0.67 \pm 0.03$.
Now we need to convert $M_V(RR)$ to $M_I(RR)$ using the mean color of
the RRab stars in our sample of $\langle$$(V-I)_0$$\rangle$$ = 0.55 \pm 0.02$.
Doing so and recomputing the distance, we arrive at 
$\langle$$(m-M)_0$$\rangle$$ = 24.61 \pm 0.07$, which agrees to within the errors
with the value determined using the V-band RR Lyrae magnitudes.

%Sarajedini et al. (2000) who used the HB magnitudes of two M33 globular clusters with 
%$[Fe/H] = -1.6$ along with the Chaboyer (1993) relation between 
%$M_V$ and $[Fe/H]$ to conclude that $\langle$$(m-M)_0$$\rangle$$ = 24.84 \pm 0.16$.
%Tiede et al. (2004) used the tip of the first ascent RGB in the I-band 
%and derived a distance of $\langle$$(m-M)_0$$\rangle$$ = 24.69 \pm 0.07$.

One could argue that, from a statistical point of view, the negative reddenings 
should not be ignored, but rather should be included in the calculation of the
distance. In this case, we find $\langle$$(m-M)_0$$\rangle$$ = 24.76 \pm 0.08$
based on the V-band RR Lyrae magnitudes and 
$\langle$$(m-M)_0$$\rangle$$ = 24.72 \pm 0.07$ from the I-band magnitudes.
Again, both of these are consistent with the mean distance modulus from the 
work of Galleti et al. (2004).

\section{Discussion}

The presence of star clusters in a kinematically hot halo has been
recognized for some time in M33. The pioneering work of Schommer et al. (1991)
and the more recent results of Chandar et al. (2002) confirm this fact. However,
the existence of a field star component to the halo has been an open question for 
some time. Studies of the halo field in M33 go back to the seminal work of
Mould \& Kristian (1986, MK86). They present a CMD of a ``halo" field  
located $\sim$20 arcmin from M33's center 
along the minor axis in the southeast direction. By 
comparing this diagram with standard giant branches, MK86 concluded that 
the mean metal abundance of the M33 halo is 
$\langle[M/H]\rangle$ = $-2.2 \pm 0.8$.  Such a metal-poor halo abundance
is in contrast to the mean metal abundance of halo clusters of 
$\langle$$[Fe/H]$$\rangle$$ = -1.27 \pm 0.11$ (Sarajedini et al. 2000).
This apparent discrepancy was resolved by Tiede, Sarajedini, \& Barker 
(2004) who reobserved
the MK86 field and constructed the deepest ground-based CMD of M33 to date.
Analysis of these data reveals that the radial metallicity gradient of the MK86
``halo" field precisely matches that of the M33 inner disk suggesting that the 
MK86 field is dominated by M33 disk stars NOT halo stars. Tiede et al. (2004)
identified problems with the color calibration of the MK86 photometry as the
source of the differences between the two studies.

The quest to identify the halo field-star population of M33 finally found 
success in the recent study by Brooks et al. (2004). They constructed a radial
stellar density profile out to 1 degree from the center of M33 corresponding
to a projected distance of about 16 kpc. The slope of their 
density profile, which includes red giant branch (RGB) stars with 20.5 $< I < $ 22.5, is
relatively shallow (--1.46 $\pm$ 0.14 in the Log $\sigma$ $\sim$ Log R plane),
suggesting the presence of a halo population.  Interestingly, the peak 
metallicity of the halo as determined by Brooks et al. (2004) is 
$[Fe/H] = -1.24 \pm 0.04$, which is gratifyingly close to the mean abundance
Sarajedini et al. (2000) derive for the halo globular clusters of 
$\langle$$[Fe/H]$$\rangle$$ = -1.27 \pm 0.11$. 

With the discovery and characterization of RR Lyraes in M33 described herein,
we have again shown that the halo field star population of M33 is non-negligible. The
peak metal abundance of these putative halo RR Lyraes is consistent with that
of the halo globular clusters presented by Sarajedini et al. (2000). In addition,
the presence of these RR Lyraes indicates that the halo field stars of M33 possess a significant
old population ($\gea$10 Gyr), which along with the diversity of HB types seen
among the halo globular clusters (Sarajedini et al. 2000), further reinforces the   
idea that the halo of M33 formed over a time period of
5 to 7 Gyr as first suggested by Sarajedini et al. (1998). This is in stark contrast
to the Milky Way halo which is mostly old and formed over only a few Gyr.
The presence of 
old ($\gea$10 Gyr), metal-rich RR Lyrae stars presumably in the disk of M33
is also very interesting and deserving of further study.

\section{Conclusions}

We present HST/ACS observations of two fields in the Local Group spiral galaxy
M33. The time baseline of our photometry from these observations spans 
$\sim$2.2 days allowing us to identify RR Lyraes in M33 from space-based
data for the first time. From the 
properties of 64 ab-type RR Lyraes, we draw the following conclusions.

\noindent 1) The minimum light colors of these stars provide a means by which
to calculate reddenings. The distribution of reddenings suggests the presence
of RR Lyraes in the halo and disk of M33. 

\noindent 2) The periods of the RR Lyraes provide a means by which to calculate
metallicities. The distribution of metallicities also suggests the presence 
of RR Lyraes in the halo and disk.

\noindent 3) Thus, both the field-halo and disk of M33 possess an old stellar
population with an age of $\gea$10 Gyr.

\noindent 4) Based on the mean V and I band magnitudes of these stars,
along with period-based metallicities, and the Chaboyer (1993) relation between
RR Lyrae metallicity and absolute magnitude, we calculate a mean distance modulus 
of $\langle$$(m-M)_0$$\rangle$$ = 24.67 \pm 0.08$ placing M33 $\sim$70 kpc
further away than M31.

\acknowledgments

We are grateful to Andy Layden for providing his suite of software for
performing the light curve fitting as well as his useful advice in the area of
RR Lyrae properties. A.S. and M.K.B. received support for this work (proposal number 
GO-9873) from  NASA through a 
grant from the Space Telescope Science Institute 
which is operated by the Association of Universities 
for Research in Astronomy, Incorporated, under NASA contract NAS5-26555.
D.G. gratefully acknowledges support from the Chilean {\sl Centro de 
Astrof\'\i  sica} FONDAP No. 15010003.

\clearpage

\begin{deluxetable}{lccccc}
\tablecaption{Observing Log}
\tablewidth{0pt}
\tablehead{
   \colhead{Field}
  &\colhead{RA  (2000)}
  &\colhead{Dec}
  &\colhead{PA V3\tablenotemark{a}}
  &\colhead{Filter}
  &\colhead{Exp Time}
}
\startdata
U49 & 1h 33m 40s & +30$^o$ 47' 59'' &  316.59 & F606W & 1 x 2494s,     3 x 2640s\\
         &                       &                               &   & F814W & 2 x 2494s,     6 x 2640s\\
         &                       &                               &   &               &                  \\
M9   & 1h 34m 30s & +30$^o$ 38' 13''  &  239.45 & F606W & 1 x 2494s,     3 x 2640s\\
         &                       &                                &  & F814W & 2 x 2494s,     6 x 2640s \\
\enddata
\tablenotetext{a}{Position angle of V3-axis of HST in degrees.}
\end{deluxetable}

\begin{deluxetable}{lcccccccc}
\tablecaption{RR Lyrae Properties}
\tablewidth{0pt}
\tabletypesize{\scriptsize}
\tablehead{
   \colhead{Name}
  &\colhead{RA  (2000)}
  &\colhead{Dec}
  &\colhead{$\langle$$V$$\rangle$}
  &\colhead{$\langle$$I$$\rangle$}
  &\colhead{$\langle$$(V-I)$$\rangle$}
  &\colhead{Period (days)}
  &\colhead{Amp. I}
  &\colhead{Type}
}
\startdata
               &                     &                     &         &     U49 - WFC1        &            &  &  & \\
  60242 & 1 33 43.08 & 30 48 33.3 & 25.477 & 24.721 & 0.825 & 0.640 & 0.557 & ab \\        
  72044 & 1 33 43.22 & 30 48  6.2 & 25.036 & 24.823 & 0.202 & 0.681 & 0.569 & c  \\        
  83628 & 1 33 36.99 & 30 48  5.0 & 25.371 & 24.959 & 0.385 & 0.329 & 0.849 & ab  \\        
  88085 & 1 33 40.72 & 30 47 46.9 & 25.406 & 25.010 & 0.431 & 0.336 & 0.912 & ab  \\        
  88219 & 1 33 45.58 & 30 47 55.2 & 25.113 & 24.879 & 0.275 & 0.501 & 0.931 & c  \\        
  88263 & 1 33 41.13 & 30 47 25.0 & 25.732 & 25.035 & 0.738 & 0.339 & 0.836 & ab  \\        
  95931 & 1 33 42.28 & 30 48 49.0 & 25.313 & 25.093 & 0.419 & 0.490 & 0.701 & ab  \\        
 100315 & 1 33 34.10 & 30 47 15.1 & 26.033 & 25.055 & 0.960 & 0.552 & 0.705 & ab  \\        
 102459 & 1 33 39.08 & 30 48 47.2 & 25.725 & 25.122 & 0.585 & 0.512 & 0.691 & c  \\        
 102996 & 1 33 41.05 & 30 47 17.6 & 26.160 & 25.133 & 1.084 & 0.611 & 0.622 & ab  \\        
 103812 & 1 33 34.27 & 30 47 45.6 & 26.056 & 25.079 & 0.970 & 0.536 & 0.429 & ab  \\        
 104521 & 1 33 38.15 & 30 46 32.9 & 26.109 & 25.133 & 0.964 & 0.553 & 0.525 & ab  \\        
 109540 & 1 33 40.62 & 30 48 31.0 & 25.789 & 25.292 & 0.466 & 0.526 & 0.740 & c  \\        
 114242 & 1 33 35.91 & 30 48  5.2 & 25.994 & 25.241 & 0.781 & 0.405 & 0.676 & ab  \\        
 116063 & 1 33 48.45 & 30 48 39.0 & 26.142 & 25.287 & 0.885 & 0.537 & 0.505 & ab  \\        
 116629 & 1 33 41.52 & 30 48 59.9 & 25.994 & 25.336 & 0.654 & 0.377 & 0.573 & ab  \\        
 119179 & 1 33 40.11 & 30 48 22.5 & 26.191 & 25.324 & 0.839 & 0.429 & 0.877 & ab  \\        
 120673 & 1 33 35.50 & 30 47 23.5 & 25.866 & 25.421 & 0.498 & 0.417 & 0.882 & ab  \\        
 121343 & 1 33 43.27 & 30 47 26.3 & 26.255 & 25.259 & 1.008 & 0.570 & 0.689 & ab  \\        
 131455 & 1 33 38.28 & 30 48  3.6 & 26.667 & 25.426 & 1.212 & 0.600 & 0.867 & ab  \\        
 132610 & 1 33 40.72 & 30 47 37.6 & 26.386 & 25.371 & 1.087 & 0.546 & 0.651 & ab  \\        
 132766 & 1 33 44.53 & 30 49 22.3 & 26.438 & 25.355 & 1.081 & 0.333 & 0.332 & c  \\        
        &            &            &        &        &       &       &       & \\        
        &            &            &        &    U49 - WFC2    &       &       &      &  \\        
  61160 & 1 33 32.58 & 30 49 35.2 & 25.794 & 24.860 & 0.955 & 0.595 & 0.617 & ab  \\        
  65431 & 1 33 38.14 & 30 50 19.2 & 26.075 & 25.080 & 0.985 & 0.530 & 0.775 & ab  \\        
  66329 & 1 33 36.15 & 30 49 19.6 & 25.712 & 25.087 & 0.632 & 0.592 & 0.845 & ab  \\        
  68673 & 1 33 33.46 & 30 48 35.4 & 26.359 & 25.091 & 1.258 & 0.536 & 0.750 & ab  \\        
  76675 & 1 33 36.07 & 30 49 36.2 & 26.010 & 25.221 & 0.791 & 0.520 & 0.813 & ab  \\        
  77426 & 1 33 31.66 & 30 48 57.4 & 26.064 & 25.052 & 0.986 & 0.498 & 0.783 & ab  \\        
  85243 & 1 33 33.22 & 30 48  3.8 & 26.196 & 25.360 & 0.840 & 0.560 & 0.738 & ab  \\        
  95229 & 1 33 29.96 & 30 48 21.0 & 26.275 & 25.511 & 0.749 & 0.250 & 0.623 & c  \\        
  97246 & 1 33 31.52 & 30 48 12.3 & 26.159 & 25.596 & 0.557 & 0.514 & 0.587 & ab  \\        
  97903 & 1 33 37.68 & 30 49 28.9 & 26.574 & 25.431 & 1.143 & 0.393 & 0.489 & ab  \\        
  99949 & 1 33 35.60 & 30 50 29.9 & 26.552 & 25.546 & 1.006 & 0.330 & 0.374 & ab  \\        
 102256 & 1 33 38.61 & 30 49 11.1 & 26.404 & 25.507 & 0.906 & 0.407 & 0.608 & ab  \\        
 102530 & 1 33 39.93 & 30 49 14.6 & 26.312 & 25.696 & 0.614 & 0.492 & 0.272 & c  \\        
 103379 & 1 33 35.23 & 30 49 22.4 & 26.249 & 25.575 & 0.798 & 0.440 & 0.839 & ab  \\        
        &            &            &        &        &       &       &       & \\        
        &            &            &        &    M9 - WFC1    &       &       &       & \\        
  21307 & 1 34 30.25 & 30 38 12.6 & 24.680 & 24.347 & 0.340 & 0.333 & 0.679 & ab  \\        
  43952 & 1 34 25.01 & 30 39  6.1 & 25.210 & 24.759 & 0.462 & 0.559 & 0.657 & ab  \\        
  45819 & 1 34 30.89 & 30 39  7.5 & 25.202 & 24.809 & 0.415 & 0.571 & 0.747 & ab  \\        
  49806 & 1 34 27.67 & 30 39 19.9 & 25.328 & 24.804 & 0.543 & 0.559 & 0.754 & ab  \\        
  50684 & 1 34 29.87 & 30 38 19.3 & 25.311 & 24.815 & 0.562 & 0.368 & 0.664 & ab  \\        
  53358 & 1 34 28.80 & 30 38 22.3 & 25.333 & 24.940 & 0.418 & 0.319 & 0.538 & ab  \\        
  55610 & 1 34 33.87 & 30 38 12.7 & 25.551 & 25.001 & 0.574 & 0.520 & 0.751 & ab  \\        
  57000 & 1 34 33.94 & 30 37 58.9 & 25.741 & 25.004 & 0.724 & 0.409 & 0.838 & ab  \\        
  57039 & 1 34 31.84 & 30 37 20.7 & 25.678 & 24.990 & 0.713 & 0.522 & 0.822 & ab  \\        
  57707 & 1 34 28.36 & 30 37 22.1 & 25.815 & 24.913 & 0.880 & 0.387 & 0.766 & ab  \\        
  60843 & 1 34 31.88 & 30 37 42.6 & 25.774 & 25.050 & 0.697 & 0.589 & 0.697 & c  \\        
  67510 & 1 34 29.86 & 30 37 41.8 & 25.662 & 25.212 & 0.518 & 0.518 & 0.791 & ab  \\        
  73401 & 1 34 32.58 & 30 37 33.0 & 26.119 & 25.393 & 0.726 & 0.322 & 0.572 & ab  \\        
        &            &            &        &        &       &       &       & \\        
        &            &            &        &  M9 - WFC2      &       &       &      &  \\        
  59370 & 1 34 20.53 & 30 38 46.7 & 25.134 & 24.740 & 0.404 & 0.321 & 0.616 & ab  \\        
  63575 & 1 34 19.00 & 30 37 49.5 & 25.402 & 24.996 & 0.422 & 0.496 & 0.902 & ab  \\        
  68851 & 1 34 20.72 & 30 36 51.9 & 25.194 & 24.865 & 0.364 & 0.512 & 0.738 & ab  \\        
  69612 & 1 34 20.71 & 30 36 44.5 & 25.382 & 24.859 & 0.555 & 0.496 & 0.564 & ab  \\        
  70023 & 1 34 22.02 & 30 38 54.3 & 25.345 & 24.907 & 0.473 & 0.333 & 0.837 & ab  \\        
  72301 & 1 34 24.65 & 30 36 25.9 & 25.343 & 24.861 & 0.513 & 0.356 & 0.601 & ab  \\        
  73898 & 1 34 19.68 & 30 37 33.1 & 25.329 & 24.832 & 0.518 & 0.529 & 0.736 & ab  \\        
  76410 & 1 34 19.59 & 30 37 14.7 & 25.657 & 24.824 & 0.827 & 0.570 & 0.698 & ab  \\        
  77532 & 1 34 22.47 & 30 37 23.3 & 25.405 & 24.896 & 0.535 & 0.320 & 0.744 & ab  \\        
  82703 & 1 34 25.74 & 30 37 53.4 & 25.530 & 25.007 & 0.546 & 0.505 & 0.739 & ab  \\        
  85311 & 1 34 26.19 & 30 37 45.4 & 25.727 & 24.848 & 0.859 & 0.546 & 0.818 & ab  \\        
  86029 & 1 34 22.22 & 30 37 55.9 & 25.816 & 24.893 & 0.906 & 0.555 & 0.701 & ab  \\        
  87001 & 1 34 23.48 & 30 38  2.6 & 25.642 & 25.060 & 0.592 & 0.463 & 0.711 & ab  \\        
  88677 & 1 34 24.48 & 30 35 55.3 & 25.791 & 24.901 & 0.864 & 0.529 & 0.749 & ab  \\        
  92546 & 1 34 26.07 & 30 36 53.9 & 25.575 & 24.944 & 0.615 & 0.395 & 0.771 & ab  \\        
  94025 & 1 34 20.39 & 30 37 12.0 & 25.962 & 25.015 & 0.934 & 0.595 & 0.644 & ab  \\        
  94890 & 1 34 22.02 & 30 37 38.9 & 25.861 & 25.077 & 0.773 & 0.552 & 0.696 & ab  \\        
  95126 & 1 34 28.83 & 30 36 55.2 & 25.970 & 25.004 & 0.965 & 0.541 & 0.701 & ab  \\        
  97250 & 1 34 29.44 & 30 36 42.0 & 25.695 & 25.247 & 0.466 & 0.420 & 0.875 & ab  \\        
 109239 & 1 34 22.33 & 30 37 28.4 & 25.431 & 25.356 & 0.092 & 0.539 & 0.524 & ab  \\        
 112403 & 1 34 23.65 & 30 37 12.7 & 25.883 & 25.607 & 0.269 & 0.396 & 0.479 & ab  \\        
 119403 & 1 34 20.41 & 30 38 29.8 & 26.641 & 25.388 & 1.238 & 0.549 & 0.796 & ab  \\        
 125762 & 1 34 23.17 & 30 36  4.7 & 26.006 & 25.532 & 0.463 & 0.484 & 0.554 & ab  \\        
   \enddata
\end{deluxetable}

%% the file, uncomment the \plotone, \plottwo, and
%% If you need a layout that cannot be achieved with \plotone or
%% \plottwo, you can invoke the graphicx package directly with the
%% \includegraphics command or use \plotfiddle. For more information,
%% please see the tutorial on "Using Electronic Art with AASTeX" in the
%% documentation section at the AASTeX Web site,
%% http://www.journals.uchicago.edu/AAS/AASTeX.
%%
%% The examples below also include sample markup for submission of
%% supplemental electronic materials. As always, be sure to check
%% the instructions to authors for the journal you are submitting to
%% for specific submissions guidelines as they vary from

%% This example uses \plotone to include an EPS file scaled to
%% 80% of its natural size with \epsscale. Its caption
%% has been written to indicate that additional figure parts will be
%% available in the electronic journal.

\clearpage
\begin{figure}
%Fig. 1
\epsscale{1.0}
%\plotone{f1.ps}
%\plotone{finder_chart.ps}
\caption{The location of our observed ACS fields overplotted on
the DSS-II image of M33. The field is approximately 20 arcmin a side;
North is up and east is to the left.}
\end{figure}

\begin{figure}
%Fig. 2
\epsscale{1.0}
%\plotone{f2.ps}
%\plotone{u49wfc1_lc1.ps}
\caption{Light curves for candidate RR Lyraes in the U49 WFC1 field. 
The open circles are the I-band data while the filled circles are the V-band.}
\end{figure}

\begin{figure}
%Fig. 3
\epsscale{1.0}
%\plotone{f3.ps}
%\plotone{u49wfc1_lc2.ps}
\caption{Same as Fig. 2.}
\end{figure}

\begin{figure}
%Fig. 4
\epsscale{1.0}
%\plotone{f4.ps}
%\plotone{u49wfc1_lc3.ps}
\caption{Same as Fig. 2.}
\end{figure}

\begin{figure}
%Fig. 5
\epsscale{1.0}
%\plotone{f5.ps}
%\plotone{u49wfc2_lc1.ps}
\caption{Light curves for candidate RR Lyraes in the U49 WFC2 field. 
The open circles are the I-band data while the filled circles are the V-band.}
\end{figure}

\begin{figure}
%Fig. 6
\epsscale{1.0}
%\plotone{f6.ps}
%\plotone{u49wfc2_lc2.ps}
\caption{Same as Fig. 5.}
\end{figure}

\begin{figure}
%Fig. 7
\epsscale{1.0}
%\plotone{f7.ps}
%\plotone{m9wfc1_lc1.ps}
\caption{Light curves for candidate RR Lyraes in the M9 WFC1 field. 
The open circles are the I-band data while the filled circles are the V-band.}
\end{figure}

\begin{figure}
%Fig. 8
\epsscale{1.0}
%\plotone{f8.ps}
%\plotone{m9wfc1_lc2.ps}
\caption{Same as Fig. 7.}
\end{figure}

\begin{figure}
%Fig. 9
\epsscale{1.0}
%\plotone{f9.ps}
%\plotone{m9wfc2_lc1.ps}
\caption{Light curves for candidate RR Lyraes in the M9 WFC2 field. 
The open circles are the I-band data while the filled circles are the V-band.}
\end{figure}

\begin{figure}
%Fig. 10
\epsscale{1.0}
%\plotone{f10.ps}
%\plotone{m9wfc2_lc2.ps}
\caption{Same as Fig. 9.}
\end{figure}

\begin{figure}
%Fig. 11
\epsscale{1.0}
%\plotone{f11.ps}
%\plotone{m9wfc2_lc3.ps}
\caption{Same as Fig. 9.}
\end{figure}

\begin{figure}
%Fig. 12
\epsscale{1.0}
%\plotone{f12small.ps}
%\plotone{contourV_new_jpg.ps}
\caption{The color-magnitude diagram of the M9-WFC1 field showing the
locations of the RR Lyrae stars identified in this paper. The reddening vector 
for E(V--I) = 0.3 is also shown.}
\end{figure}

\begin{figure}
%Fig. 13
\epsscale{1.0}
%\plotone{f13.ps}
%\plotone{LogP_Amp.eps}
\caption{The period-amplitude diagram for the RR Lyraes in M33. The open and
filled circles are ab-type and c-type variables, respectively. The solid lines are
the loci for Oosterhoff I (left) and II (right) globular clusters from Clement (2000).
These have been converted from V-band amplitudes to I-band by dividing
them by 1.6 as per the results of Liu \& Janes (1990).}
\end{figure}

\begin{figure}
%Fig. 14
\epsscale{1.0}
%\plotone{f14.ps}
%\plotone{LogP_dist.eps}
\caption{The distribution of periods for the ab-type RR Lyraes in the present work (solid 
line). Analogous distributions for the M31 halo (Brown et al. 2004) and the globular
cluster M3 (Brown et al. 2004) are shown as the dashed and dotted lines,
respectively. These have been scaled to have the same number of RR Lyraes as the M33
distribution.}
\end{figure}

\begin{figure}
%Fig. 15
\epsscale{1.0}
%\plotone{f15.ps}
%\plotone{red_hist.eps}
\caption{The distribution of $E(V-I)$ reddenings determined for the 63 RRab stars
with adequate light curves using the method of Sturch (1966). The dotted lines are
Gaussian fits to the negative and positive portions of the reddening distribution.}
\end{figure}

\begin{figure}
%Fig. 16
\epsscale{1.0}
%\plotone{f16.ps}
%\plotone{RRLs_ACL.eps}
\caption{A plot of the metal abundance of Galactic RRab stars in the Layden (2005, private
communication) database as a function of the logarithm of the period in days (Log P). The
dashed line is the best-fit relation determined  by performing a least-squares
fit using Log P as the independent variable and then with [Fe/H] as the independent
variable and combining the results (see text). The root-mean-square deviation of the
points from the fit is 0.45 dex.}
\end{figure}

\begin{figure}
%Fig. 17
\epsscale{0.9}
%\plotone{f17.ps}
%\plotone{met_hist.eps}
\caption{The distribution of metallicities determined from  the period of each M33
RRab star. The dashed line is the generalized histogram of metallicity values
assuming an error of 0.45 dex for each value. The solid line is the sum
of two Gaussian distributions, one with a peak at $[Fe/H] = -1.3$ and the other 
with a peak at $[Fe/H] = -0.7$, fitted to the generalized histogram.}
\end{figure}

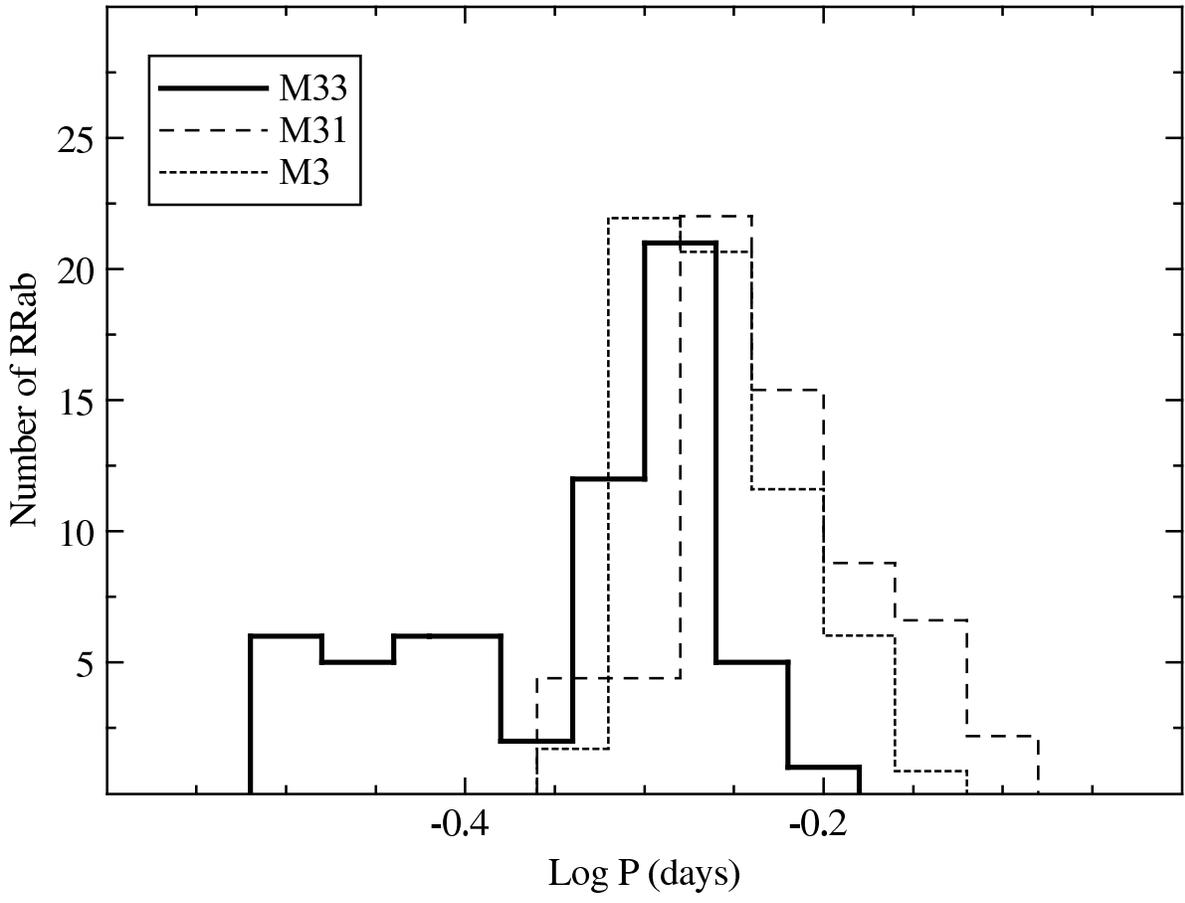
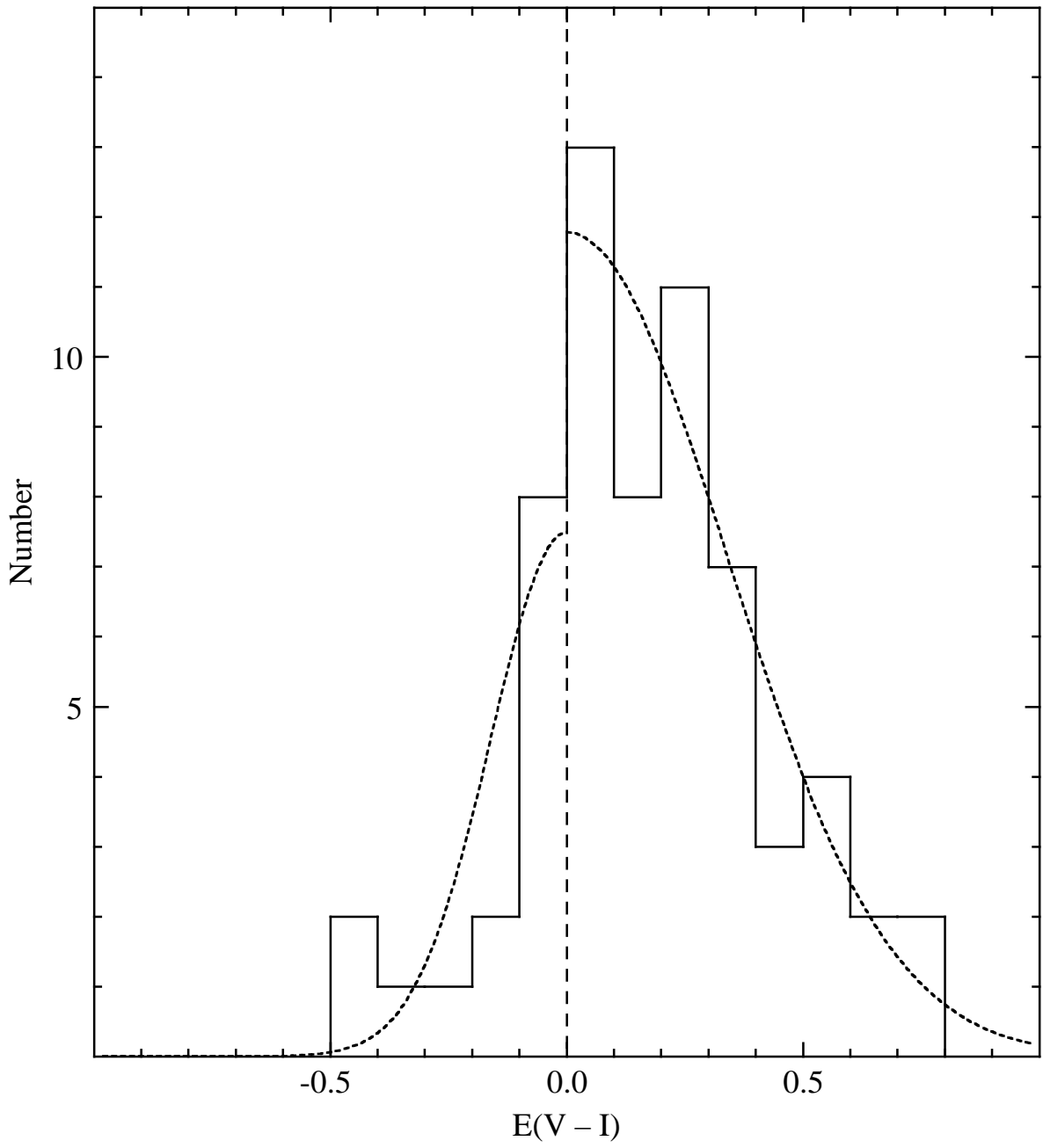
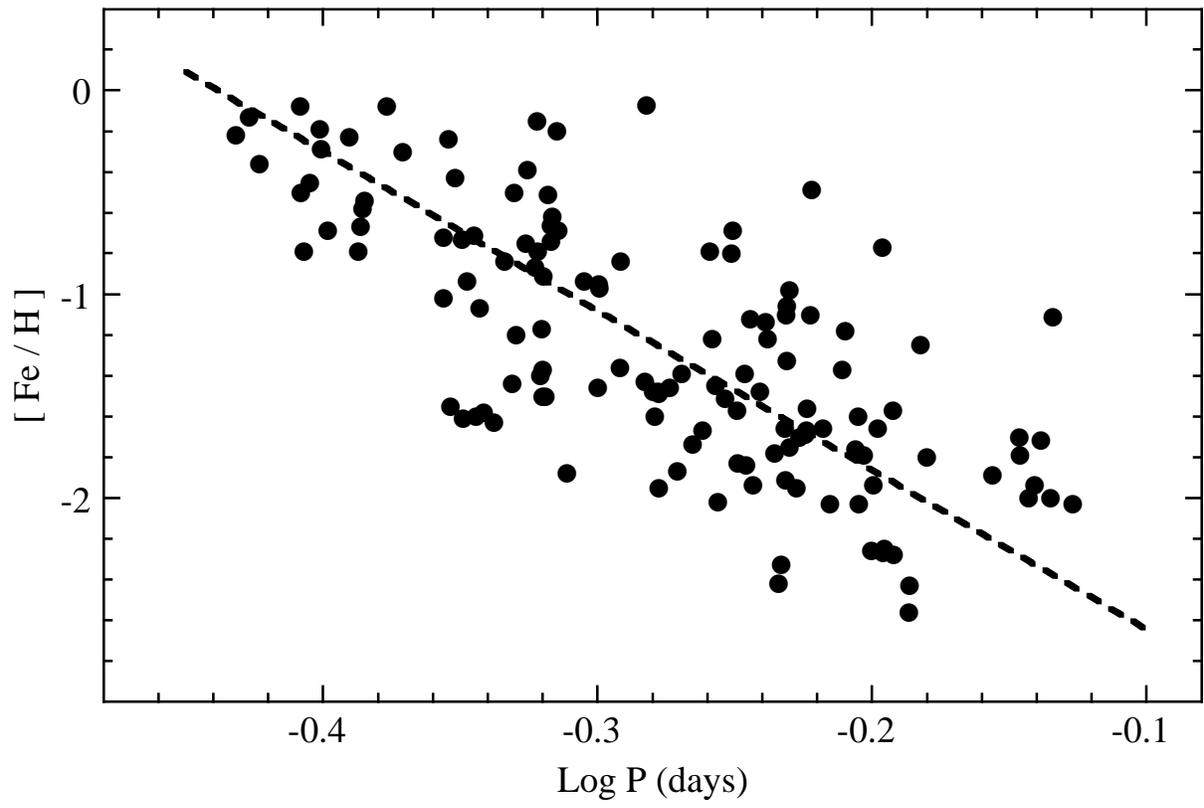
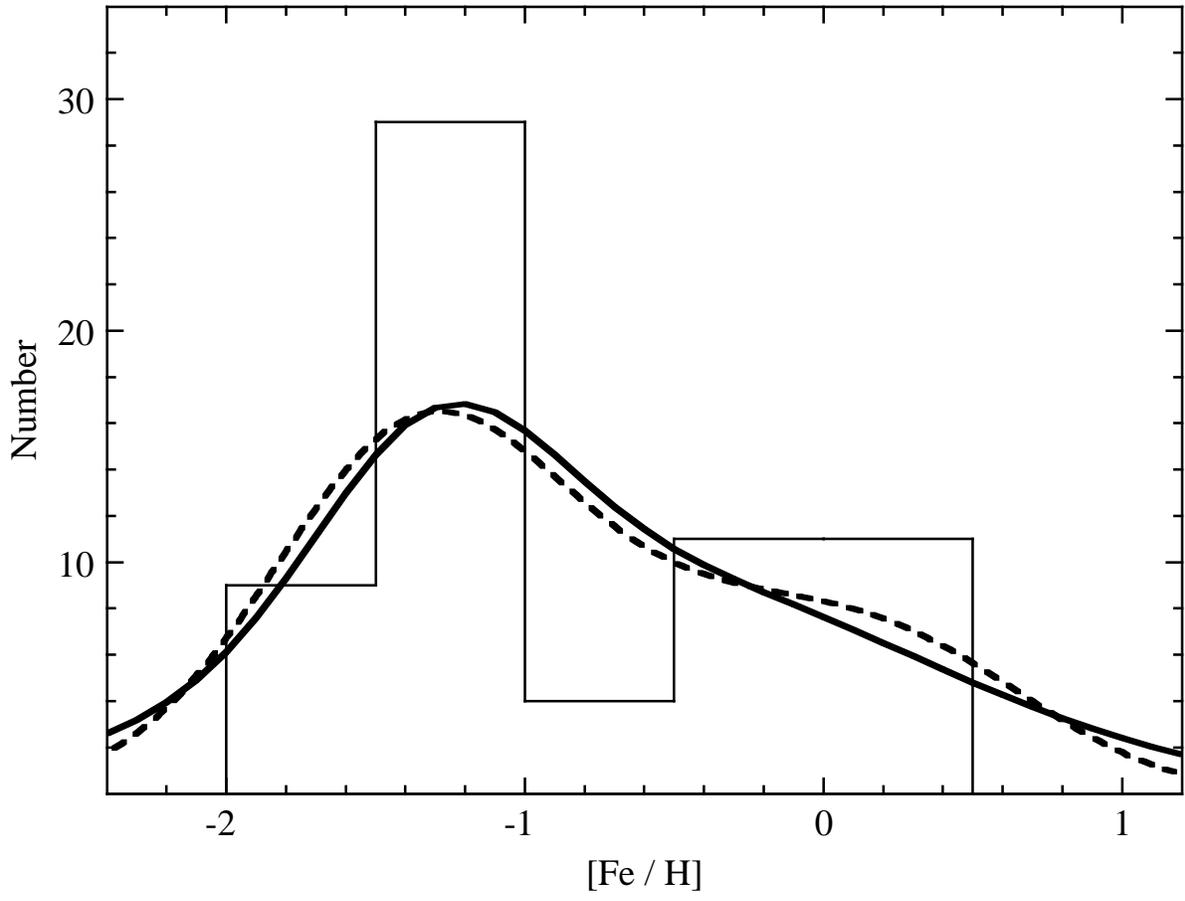
\end{document}